\documentclass[letterpaper,conference]{IEEEtran}
\usepackage{setspace}

\usepackage{amsmath}
\usepackage{amssymb}
\usepackage[hyphens]{url}
\usepackage{hyperref}
\hypersetup{colorlinks=true,breaklinks=true}
\usepackage[pdftex]{graphicx}
\usepackage{algorithm}
\usepackage[noend]{algpseudocode}
\usepackage{cite}
\usepackage{amsfonts}
\usepackage{textcomp}
\usepackage{xcolor}
\usepackage{afterpage}
\usepackage[paper=letterpaper,top=1in,bottom=0.75in,right=0.75in,left=0.75in]{geometry}
\usepackage[font=normalsize]{caption}

\def\BibTeX{{\rm B\kern-.05em{\sc i\kern-.025em b}\kern-.08em
    T\kern-.1667em\lower.7ex\hbox{E}\kern-.125emX}}

\newtheorem{theorem}{Theorem}

\newtheorem{remark}{Remark}

\title{Optimal Cruise Airspeed for Hybrid-Electric and Electric Aircraft: Applications to Air Mobility}
\author{Steven Li and Luis Rodrigues\\
Department of Electrical and Computer Engineering\\
Concordia University}

\begin{document}
\maketitle
\begin{abstract}
Electric and hybrid-electric aircraft can help our society transition towards more sustainable aviation and lower greenhouse gas (GHG) emissions. This paper provides solutions to minimize the direct operating cost (DOC) for hybrid-electric aircraft.  The solution is the positive real root of a quintic polynomial which is derived using Pontryagin’s minimum principle. By properly selecting a hybridization factor, one can also find the cruise airspeed corresponding to the minimum DOC of an electric aircraft. The optimal airspeed is integrated into the Rapidly-exploring Random Trees Star (RRT*) path planning algorithm. The minimum DOC solutions are investigated in a hybrid-electric international travel scenario and the path planning approach is applied to an electric aircraft city scenario.
\end{abstract}

\section{Introduction}
\IEEEPARstart{A}{s} the effects of climate change increase in intensity and frequency, the international community set itself to combat its leading cause: GHG emissions. Notably, the International Civil Aviation Organization (ICAO) established in 2010 a set of objectives to increase fuel efficiency and reduce emissions \cite{ICAO_clim_40}. One course of action undertaken by the aerospace industry is to develop hybrid-electric and electric aircraft. A major hurdle for electric aircraft is battery energy density which makes them better suited for urban and regional air mobility \cite{ICAO_elec_list}. In contrast, hybrid-electric aircraft were found to significantly reduce fuel consumption while maintaining the endurance of conventional aircraft \cite{hiserote2010analysis}\cite{chen2010hybrid}\cite{gibson2010potential}. With the rapid emergence of urban air mobility \cite{uamtrend2022} and the projected increase in global air traffic \cite{ICAOtrafficforecast}, it becomes imperative to quicken the integration of lower-emission aircraft into local and global fleets. 

In the mid-2000s, researchers started to investigate different control strategies to reduce fuel consumption in hybrid-electric UAVs \cite{Harmon_2005} \cite{Harmon_2006}. Following ICAO's goals statement in 2010, research and development of hybrid-electric aircraft increased drastically. The authors of \cite{Pornet_2014} provided two metrics, a cost-specific air range and a cost index relating time to energy cost, that can be used to minimize the direct operating cost (DOC) for a hybrid-electric aircraft. Falck and Gray \cite{falck2016parallel} tackled the trajectory optimization problem for hybrid-electric aircraft using a collocation method. The authors of \cite{bongermino2017model} used dynamic programming to find an optimal power management strategy for these aircraft. The work in \cite{donateo2017fuel} investigated the endurance of hybrid-electric aircraft by developing an on-off energy management strategy. In 2018, reference \cite{boggero2018trade} proposed the use of fuzzy logic to minimize fuel at a preliminary design level. The work in \cite{donateo2018method} used a genetic algorithm to find the engine operating points that can reduce fuel consumption. The authors of \cite{geiss2018optimized} discretized a continuous DOC optimal control problem and used \texttt{fmincon} in MATLAB to solve it. More recently,  the authors of \cite{doff2020optimal} proposed a solution to the energy management problem of hybrid-electric aircraft based on model predictive control and convex optimization while reference \cite{leite2020optimal} used dynamic programming. Other work focused on lowering operating costs through aircraft sizing and aircraft propulsion design \cite{finger2020comparative}\cite{finger2020comparison}\cite{wheeler2021electric}.

Although solutions for the minimum DOC problem of fuel and electric aircraft in cruise flight were found by \cite{Villaroel_Jan_2016} and \cite{kaptsov2018electric}, respectively, to the best of the authors' knowledge, only references \cite{Pornet_2014} and \cite{geiss2018optimized} sought to minimize the DOC for hybrid-electric aircraft.  Unfortunately, neither paper provided a systematic procedure to obtain solutions for the continuous-time problem. By contrast, this paper solves the continuous-time DOC optimal control problem and provides a quintic polynomial whose positive real root is the optimal solution that minimizes the DOC. In addition, this work defines two coefficients to trade-off between time and average energy costs and between fuel and electricity costs. It also utilizes an energy management parameter to describe the proportion of the thrust coming from electrical energy. The solutions are derived based on Pontryagin's minimum principle (PMP). Furthermore, this work implements the optimal cruise airspeed solutions in a path-planning method. The main contributions of this paper are as follows.
\begin{enumerate}
    \item A quintic polynomial whose positive real root is the optimal cruise airspeed solution of the minimum DOC problem of hybrid-electric aircraft was derived by using two trade-off coefficients and an energy management parameter.
    \item These solutions along with those for electric aircraft were combined into a path-planning algorithm to find minimum DOC paths.
\end{enumerate}

This paper is structured as follows. The hybrid-electric aircraft dynamics are presented in Section II. Then, in Section III, the minimum DOC problem is formulated and solved. In Section IV, the RRT* algorithm is briefly reviewed. Finally, simulation examples are shown for a hybrid-electric and an electric aircraft in Section V.
\newgeometry{top=0.75in,bottom=0.75in,left=0.75in,right=0.75in}
\section{Aircraft Dynamics}
For a fixed-wing hybrid-electric aircraft, we assume that:
\begin{enumerate}
    \item the electric engines' supply voltage remains relatively constant as the state of charge varies. This assumption is reasonable given that the batteries should not be fully discharged during flight \cite{shepherd1965design},
    \item the internal resistance of the battery is small and can thus be neglected and its effects can be integrated into the system through a slight reduction in the total system efficiency $\eta$,
    \item thermal effects on the batteries are neglected.
\end{enumerate}
By the definition of efficiency and Faraday's law, the electricity consumption of the aircraft can be expressed as
\begin{equation}
    \dot{Q} = -\frac{\beta Tv}{\eta U}
\end{equation}
where $Q$ is the electric charge of the batteries, $T$ is the magnitude of the aircraft's thrust force, $v$ is the airspeed, and $U$ is the supply voltage of the electric engines. Since hybrid-electric aircraft utilize two types of energy, electricity and fuel, a hybridization factor $\beta \in [0,1]$ is introduced. This coefficient describes the proportion of the thrust generated from electric engines. Given that the hybrid-electric aircraft also uses jet engines, the fuel consumption, or the change in the aircraft weight, can be described by \cite{Tewari_adv_ctrl}
\begin{equation}
    \dot{W} = -S_{fc}(1-\beta)T
\end{equation}
where $W$ is the weight and $S_{fc}$ is the thrust specific fuel consumption. Observe that when $\beta = 1$, the aircraft only uses electrical energy (electric aircraft). Conversely, when $\beta = 0$, the aircraft only consumes fuel energy (jet aircraft).

Let $r$ describe the horizontal position of the aircraft, $h$ be the altitude, $\gamma$ be the flight path angle, $\alpha$ be the angle of attack, $g$ be the gravitational acceleration, $D$ be the magnitude of the drag force, $L$ be the lift force magnitude. The dynamic model of the hybrid-electric aircraft will take into consideration the following assumptions.
\begin{enumerate}
    \item The aircraft cruises at a constant altitude. Therefore, $\dot{\gamma} = \gamma = \dot{h} = 0$, and the air density remains constant.
    \item The angle of attack of the aircraft is small. Therefore, one can approximate $\cos{\alpha} \approx 1$, and $\sin{\alpha} \approx \alpha$.
    \item The aircraft is in steady flight (i.e. $T = D$ and $L=W$) and is below the drag divergence Mach number. 
    \item The thrust perpendicular to the velocity is small compared to the weight and lift force (i.e. $T\sin{\alpha} \ll W$ and $T\sin{\alpha} \ll L$).
    \item The specific fuel consumption in cruise is assumed to be a function of the altitude $h$ (i.e. $S_f = S_f(h)$).
    \item The altitude, thrust, and speed of the aircraft will be within the flight envelope.
\end{enumerate}
Under the above assumptions, the dynamic model can be written as
\begin{equation}\label{simpdotx}
    \dot{r} = v
\end{equation}
\begin{equation}
    \dot{Q} = -\frac{\beta Dv}{\eta U} = -i
\end{equation}
\begin{equation}\label{simpdotW}
    \dot{W} = -S_{fc}(1-\beta)D = -f
\end{equation}
where $i$ is the output current from the batteries, and $f$ is the fuel consumption rate of the hybrid-electric aircraft.

Given that the aircraft is in steady flight, is below the drag divergence Mach number, and follows a drag polar curve, one can write the coefficients of lift and drag as
\begin{equation}
    C_L = \frac{2W}{\rho Sv^2}
\end{equation}
\begin{equation}
    C_D = C_{D,0} + C_{D,2}\left(\frac{2W}{\rho Sv^2}\right)^2
\end{equation}
where $\rho$ is the air density, $S$ is the wing surface area, $C_{D,0}$ is the zero-lift drag coefficient, and $C_{D,2}$ is the induced drag coefficient. The drag force can then be written as
\begin{equation}\label{D}
    D = \frac{1}{2}\rho Sv^2C_D = \frac{1}{2}C_{D,0}\rho Sv^2 + \frac{2C_{D,2}W^2}{\rho Sv^2}
\end{equation}

\section{Direct Operating Costs Minimization}
The direct operating cost (DOC) of a hybrid-electric aircraft can be written as
\begin{equation}\label{DOC}
    DOC = \int_0^{t_f}(C_t + C_i\kappa_iUi + C_f\kappa_ff)dt
\end{equation}
where $C_t$, $C_i$, and $C_f$ are the time-related cost (in units of currency per unit of time), electricity cost (in units of currency per unit of energy), and fuel cost (in units of currency per unit of energy) coefficients, respectively, $\kappa_i$ converts between two different units of energy (e.g. Joules to Kilowatt-hours), and $\kappa_f = eg^{-1}$ converts between units of energy and units of weight given $e$ as the heating value of the fuel in units of energy per unit of weight. These two conversion coefficients are used to ensure that the electricity and fuel terms in (\ref{DOC}) are in the same units.

Defining two coefficients, 
\begin{equation*}
    C_\mu = \frac{C_i + C_f}{2}, \quad C_\Delta = \frac{C_i-C_f}{2}
\end{equation*}
one can write two trade-off coefficients
\begin{equation*}
    C_I = \frac{2C_t}{C_\mu},\quad C_E = \frac{C_\Delta}{C_\mu}
\end{equation*}
given that $C_\mu > 0$, where $C_I$ describes the trade-off between the time-related costs and the average energy cost while $-1\leq C_E\leq 1$ captures a trade-off between electricity and fuel costs. 

Minimizing the DOC in (\ref{DOC}) with respect to (\ref{simpdotx})-(\ref{simpdotW}) and (\ref{D}) can be formulated as the following optimal control problem
\begin{equation}\label{OCP}
    \begin{aligned}
        J^* = \min_{v,t_f}&\int_0^{t_f}(C_I + (1+C_E)\kappa_iUi + (1-C_E)\kappa_ff)dt\\
        s.t.\quad &\dot{r} = v\\
        &\dot{W} = -S_{fc}(1-\beta)D = -f\\
        &\dot{Q} = -\frac{\beta Dv}{\eta U} = -i\\
        &r(0) = r_0,\ r(t_f) = r_f,\\
        &W(0) = W_0,\ Q(0) = Q_0\\
        &v>0
    \end{aligned}
\end{equation}
\begin{theorem}{\cite{li2021flight}}
    The optimal solution of the optimal control problem (\ref{OCP}) $v^*$ is a positive real root of the quintic equation
    \begin{equation}\label{quintic}
    \begin{aligned}
        &\frac{(1+C_E)\kappa_i\beta\rho^2S^2C_{D,0}}{\eta}v^{*^5} + \frac{\bar J^*_W(1-\beta)S_{fc}\rho^2S^2C_{D,0}}{2}v^{*^4}\\ &- C_I\rho Sv^{*^2} - \frac{4(1+C_E)\kappa_i\beta C_{D,2}W^2}{\eta}v^{*}\\ &- 6\bar J^*_W(1-\beta)S_{fc}C_{D,2}W^2 = 0
    \end{aligned}
    \end{equation}
    where
    \begin{equation}\label{JWbarchp3}
        \bar{J}^*_W = (1-C_E)\kappa_f - J^*_W
    \end{equation}
    and the time derivative of $J^*_W$ is given by
    \begin{equation}\label{dotJWchp3}
        \dot{J}^*_W = -\frac{4(1+C_E)\kappa_i\beta C_{D,2}W}{\eta\rho Sv^*} - \bar{J}^*_W(1-\beta)S_{fc}\frac{4C_{D,2}W}{\rho Sv^{*^2}}
    \end{equation}
    with final condition
    \begin{equation}
        J^*_W(t_f) = 0
    \end{equation}
\end{theorem}
\begin{IEEEproof}
The proof is detailed in \cite{li2021flight}.
\end{IEEEproof}

Given that no general analytical methods exist for solving quintic equations, one needs to employ numerical methods to determine the optimal cruise airspeed of a hybrid-electric aircraft.

\begin{remark}
    Observe that for an electric aircraft, $\beta = 1$ and $W$ is constant. Therefore, (\ref{quintic}) reduces to a quartic equation for $v^*>0$ that can be solved analytically as shown in \cite{kaptsov2018electric}.
\end{remark}

\section{Trajectory Planning with RRT*}
\begin{figure}[h]
    \centering
    \includegraphics[scale=0.26]{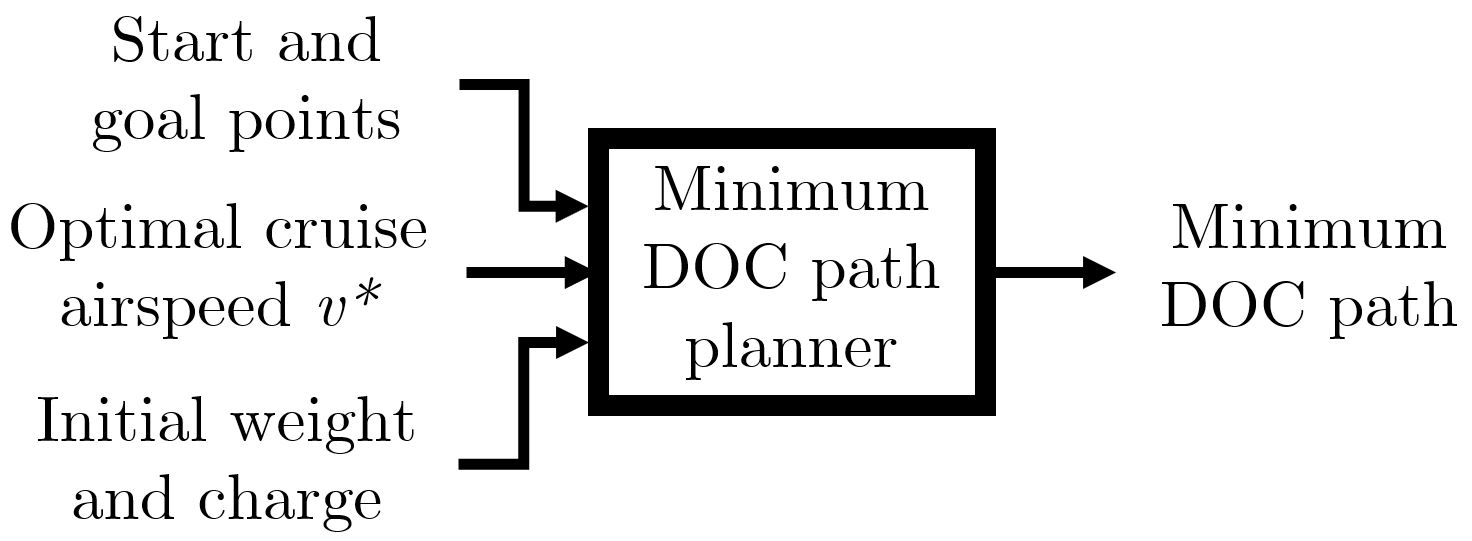}
    \caption{Block diagram of the minimum DOC path planning process using optimal cruise airspeed solutions.}
    \label{fig:pathplanningDOC}
\end{figure}

In the context of air mobility, especially in urban areas, path-planning algorithms are important for aircraft to navigate within complex environments. Incorporating cost-minimization capabilities for electric and hybrid-electric aircraft in these algorithms can assist with the sustainability of air mobility. To achieve this, the optimal cruise airspeed obtained from theorem 1 can be applied to a path planning algorithm. This process is summarized in Figure \ref{fig:pathplanningDOC}.

\begin{figure}[h]
    \centering
    \includegraphics[scale=0.5967]{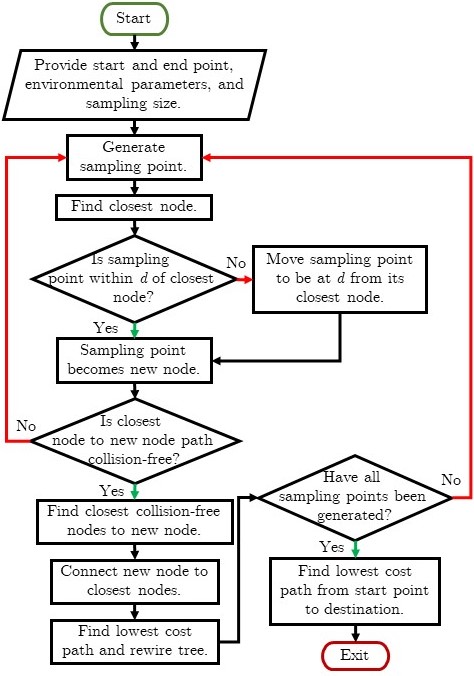}
    \caption{Process flowchart of the Rapidly-exploring Random Trees Star (RRT*) algorithm.}
    \label{fig:flowchart}
\end{figure}

Numerous path-planning algorithms have been developed to tackle the challenges of navigation in environments with obstacles. The results from this work can be applied to path planning methods whose objective is to minimize a given cost. One such algorithm is a sampling-based approach known as Rapidly-exploring Random Trees* (RRT*) \cite{karaman2011sampling}. This path-planning algorithm asymptotically converges to an optimal solution as the number of iterations approaches infinity. RRT* was selected for path planning in this work because it offers scalability to higher dimensions \cite{bialkowski2011massively} and it has been applied in the handling of many complex obstacle-avoidance problems \cite{kang2021smooth}\cite{chen2020fuzzy}. The RRT* algorithm samples the configuration space and is summarized in the flowchart of Figure \ref{fig:flowchart}.

The number of iterations in the RRT* algorithm is decided by the number of sampling points provided at the start of the procedure. The RRT* path planner generates a tree-like structure of possible paths from a start position to a goal position and iteratively improves the solution by rewiring the tree to reach the goal in the shortest possible distance. This distance can be abstracted to the notion of a cost. In this work, the shortest distance is replaced by the minimum direct operating cost of a hybrid-electric aircraft. The in-depth details and derivations of the RRT* algorithm can be found in \cite{karaman2011sampling}. 

\section{Simulations Examples}
Simulation scenarios in international travel and urban air mobility are presented where the optimal cruise airspeed is found using theorem 1. The airspeed is computed using a shooting method and a polynomial root solver in MATLAB. The first scenario investigates the long-range applicability of hybrid-electric aircraft while validating the results of theorem 1. Since this scenario is assumed to be obstacle-free, no path-planning algorithm is used. The second scenario requires navigation in an urban area with obstacles such as buildings. Therefore, RRT* will be used to plan the aircraft path.  

For both scenarios, a gravitational acceleration value $g$ of 9.8 m/s$^2$, an electricity cost coefficient $C_i$ of 0.06 USD/kWh, and an electricity conversion factor $\kappa_i$ of 1/3.6$\times10^6$ kWh/Joules are used. The simulations are performed on a laptop equipped with 16 GB of RAM and an Intel Core i7-7700HQ 2.80 GHz CPU.

\subsection{International Travel Hybrid-Electric Aircraft Scenario}
For the international travel scenario, the Airbus E-Fan X aircraft model is used and its parameters are listed in Table \ref{tab:EFanXparams}. Although the aircraft's development has been discontinued, the E-Fan X can still provide a benchmark for studying the results proposed in this paper. The E-Fan X's structure is based on the British Aerospace 146 aircraft \cite{efanx}. A hybrid-electric aircraft was selected for this scenario because it is more suitable for longer-range commercial flights \cite{ICAO_elec_list}. The aircraft starts at a location close to Pierre Elliott Trudeau International Airport in Montreal, Canada and cruises to a location near John F. Kennedy International Airport in New York City, USA. The aircraft flies at a constant altitude of 10 km above sea level. At this altitude, the air density $\rho$ is around 0.4135 kg/m$^3$ \cite{engrtlbxdensity}.
\begin{table}[h]
    \centering
    \caption{Airbus E-Fan X Aircraft \cite{rj100british}\cite{efanx}}
    \begin{tabular}{c c}\hline
         \textbf{Parameter}&\textbf{Value} \\\hline
         Wing surface area $S$& 77.3 m$^2$\\
         Empty weight & 25,645 kg\\
         Maximum take-off weight&44,225 kg\\
         Thrust specific fuel consumption $S_{fc}$&2.55$\times$10$^{-5}$ kg/Ns\\
         Battery output voltage $U$&3,000 V\\
         Maximum battery capacity&1,560 kWh\\
         Total electrical system efficiency $\eta$&0.9\\
         Zero-lift drag coefficient $C_{D,0}$&0.028\\
         Induced drag coefficient $C_{D,2}$&0.026\\
         Hybridization factor $\beta$&0.25\\
         Time cost coefficient $C_t$&0.5 USD/s\\
         Electricity cost coefficient $C_i$&0.06 USD/kWh\\
         Fuel cost coefficient $C_f$&0.115 USD/kWh\\\hline
    \end{tabular}
    \label{tab:EFanXparams}
\end{table}
Since the E-Fan X consumes kerosene with a heating value $e$ of approximately 11.94 kWh/kg \cite{airbpkerosene}, the fuel conversion factor is 1.22 kWh/N. Given kerosene prices of 1.12 USD/L, the fuel cost coefficient $C_f$ is chosen as 0.115 USD/kWh. Let us assume, for this scenario, a block-hour operations cost of 1,800 USD/hour, or equivalently, 0.5 USD/second. We also assume a high total electrical system efficiency $\eta$ of 0.9. The boundary conditions of this scenario are provided in Table \ref{tab:boundaryinternational}.

\begin{table}[h]
    \centering
    \caption{Simulation Conditions for the International Travel Scenario}
    \begin{tabular}{l c}\hline
         \textbf{Parameter}&\textbf{Value}\\\hline
         Initial position $r_0$& 0 km\\
         Initial weight $W_0$&430,000 N\\
         Initial charge $Q_0$&1,516,000 C\\
         Final position $r_f$& 450 km\\
         Altitude $h$&10 km\\
         Air density $\rho$&0.4135 kg/m$^3$\\\hline
    \end{tabular}
    \label{tab:boundaryinternational}
\end{table}

\begin{figure}
    \centering
    \includegraphics[scale=0.4455]{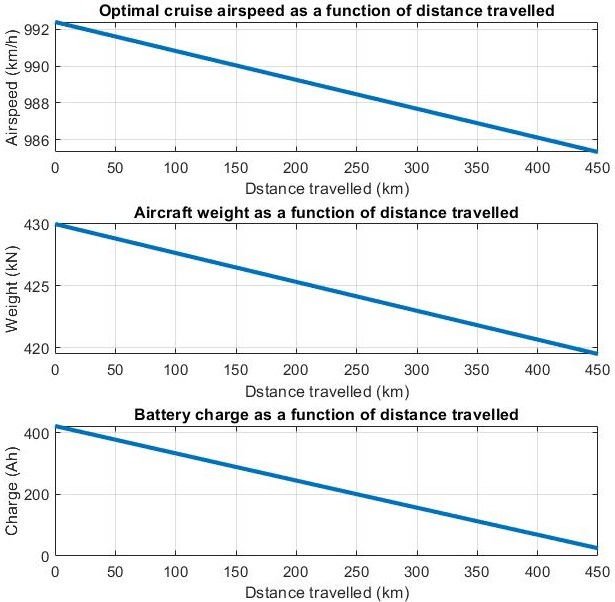}
    \caption{Optimal cruise airspeed, aircraft weight, and battery charge as a function of the distance travelled in the international travel scenario.}
    \label{fig:EFanXairspeed}
\end{figure}

For this flight, the hybrid-electric aircraft consumed 1,072 kg of kerosene and 396 Ah of charge. The duration of the flight is 1,638 seconds or a little bit over 39 minutes. The total direct operating cost of the flight is 2,362.16 USD which equates to an hourly cost of 5,191.56 USD/hour. The optimal cruise airspeed as a function of distance travelled is shown in Figure \ref{fig:EFanXairspeed}. 
\begin{figure*}
    \centering
    \includegraphics[scale=0.3765]{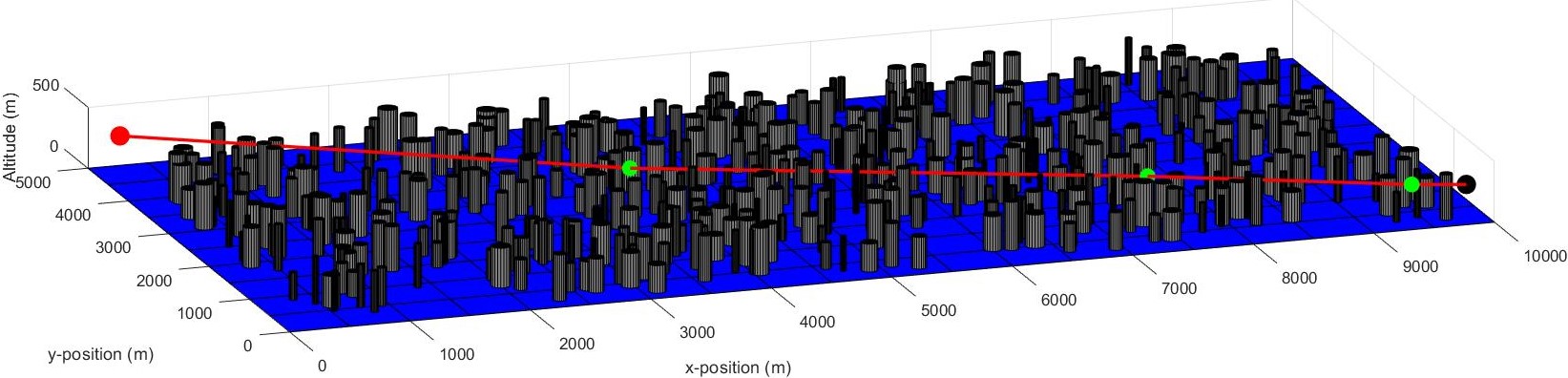}
    \caption{Minimum DOC path of the Yuneec International E430 aircraft in the city scenario. The red sphere is the starting point and the black sphere is the destination. The green spheres are waypoints computed by the RRT* algorithm along the path. The red line represents the aircraft's path. The grey cylinders are the obstacles and the blue plane is the ground plane.}
    \label{fig:pathcity}
\end{figure*}
Since the aircraft consumes fuel, the weight decreases over the span of the flight. The airspeed also decreases proportionally. Figure \ref{fig:EFanXairspeed} illustrates this proportionality.

\subsection{Urban Electric Aircraft Scenario}
For the city scenario, the Yuneec International E430 two-seater all-electric aircraft model is used. Its parameters are listed in Table \ref{tab:E430params}. An all-electric aircraft was chosen for the urban scenario as they are more suitable for shorter-range, lower-speed applications \cite{ICAO_elec_list}. In this scenario, a city with 500 buildings is randomly generated through a uniform distribution. The buildings are assumed to have 150 meters of restricted airspace directly above them and are surrounded by buffer zones. The restricted airspace for the buildings is modelled as cylindrical obstacles with radii between 20 and 80 meters and heights between 200 and 400 meters. The city is represented in a 10 km by 5 km space. 

The aircraft starts at one end of the city and flies to the other end. It navigates at a constant altitude of 300 meters. At this altitude, the air density $\rho$ is approximately 1.2 kg/m$^3$ \cite{engrtlbxdensity}.
\begin{table}[h]
    \centering
    \caption{Yuneec International E430 Aircraft Parameters \cite{greenwinge430}}
    \begin{tabular}{l c}\hline
         \textbf{Parameter}&\textbf{Value} \\\hline
         Wing surface area $S$& 11.37 m$^2$\\
         Empty weight& 302 kg\\
         Maximum take-off weight&472 kg\\
         Lift-to-drag ratio $L/D$&28\\
         Battery output voltage $U$&133.2 V\\
         Maximum battery capacity&13.32 kWh\\
         Total electrical system efficiency&0.7\\
         Zero-lift drag coefficient$^a$ $C_{D,0}$&0.035\\
         Induced drag coefficient$^a$ $C_{D,2}$&0.009\\
         Hybridization factor $\beta$&1\\
         Time cost coefficient $C_t$&0.0005 USD/s\\
         Electricity cost coefficient $C_i$&0.06 USD/kWh\\
         Fuel cost coefficient $C_f$&0.00 USD/kWh\\\hline
         \footnotesize{$^a$ Estimates}
    \end{tabular}
    \label{tab:E430params}
\end{table}
The zero-lift drag coefficient $C_{D,0}$ is chosen to be within the range of a typical general aviation aircraft \cite{roskam1997airplane}. The induced drag coefficient $C_{D,2}$ is estimated based on the lift-to-drag ratio \cite{Anderson_air_perf} as
\begin{equation}\label{LDandCD2}
\begin{aligned}
    &L/D = \frac{1}{2}\sqrt{\pi A\zeta/C_{D,0}}\\
    &C_{D,2} = 1/(\pi A\zeta)
\end{aligned}
\end{equation}
where $A$ is the aspect ratio, and $\zeta$ is the efficiency factor. Solving for the two unknowns $C_{D,2}$ and $\pi A\zeta$ in (\ref{LDandCD2}), we obtain an induced drag coefficient $C_{D,2}$ of 0.009. Let us assume a block-hour operations cost of 1.8 USD/hour, or 0.0005 USD/second and a total electrical system efficiency of 0.7. The boundary conditions of this scenario are provided in Table \ref{tab:boundarycity}. The position $r$ can be interpreted as the magnitude of a position vector $\mathbf{r}$ in the horizontal plane. Therefore, given an orientation $\psi$ with respect to the inertial reference frame, one can decompose the position vector into two coordinates
\begin{equation*}
    x = r\cos\psi,\ y = r\sin\psi
\end{equation*}
One can then define the initial and final points, with $\psi(0) = \psi_0$, and $\psi(t_f) = \psi_f$ as
\begin{align*}
    &x(0) = x_0 = r_0\cos\psi_0,\ y(0) = y_0 = r_0\sin\psi_0\\
    &x(t_f) = x_f = r_f\cos\psi_f,\ y(t_f) = y_f = r_f\sin\psi_f
\end{align*}
\begin{table}[h]
    \centering
    \caption{Simulation Conditions for the City Scenario}
    \begin{tabular}{c c}\hline
         \textbf{Parameter}&\textbf{Value}\\\hline
         Initial position $(x_0,y_0)$&(200,4800) m\\
         Initial weight $W_0$&4,600 N\\
         Initial charge $Q_0$&360,000 C\\
         Final position $(x_f,y_f)$&(9800,100) m\\
         Altitude $h$&300 m\\
         Air density $\rho$&1.2 kg/m$^3$\\\hline
    \end{tabular}
    \label{tab:boundarycity}
\end{table}
\begin{figure}[H]
    \centering
    \includegraphics[scale=0.47]{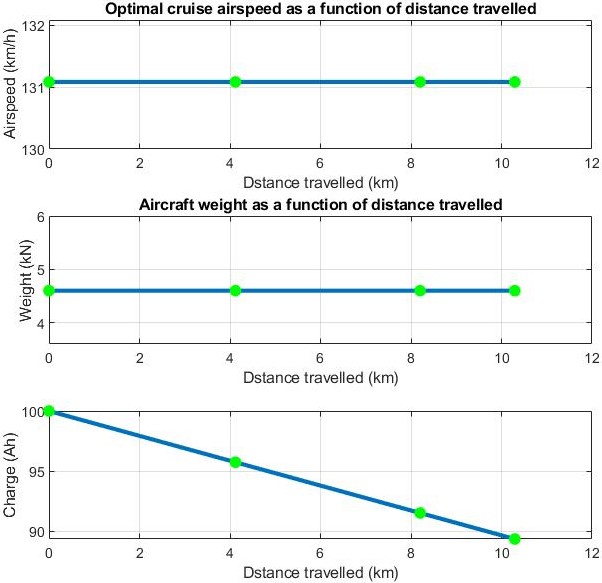}
    \caption{Optimal cruise airspeed, aircraft weight, and battery charge as a function of the distance travelled in the urban scenario. The green circles are waypoints generated by RRT*.}
    \label{fig:E430airspeed}
\end{figure}
The RRT* algorithm described in Section IV is implemented in MATLAB. A total of 250 points are sampled through a Gaussian distribution based on \cite{lichocki2020gaussian}. The lowest DOC path, shown in Figure \ref{fig:pathcity}, was found in 335 seconds. 

Since the Yuneec International E430 is an all-electric aircraft, its weight does not change as shown in Figure \ref{fig:E430airspeed}. As a result, the optimal cruise airspeed also remains constant at a value of 130 km/h. The total electricity consumed is 10.7 Ah over a distance of 10.294 km which is well within the capacities of the aircraft batteries. This result shows that smaller electric aircraft can be feasible options for urban air mobility due to their lower weight and speeds. The total direct operating cost of the flight is 0.23 USD. The aircraft flew for a duration of 297 seconds. The hourly direct operating cost of the aircraft is 2.79 USD/hour. If the aircraft consumed kerosene instead of electricity, with a $S_{fc} = 1.1\times10^{-5}$ kg/Ns, and followed the same path, the DOC of the flight would rise to 1.48 USD which is an over six-fold increase. Therefore, it is cheaper to fly an electric aircraft in this scenario.

\section{Conclusions}
To help the transition to more sustainable aviation, this paper derived a quintic polynomial whose positive real root is the optimal cruise airspeed that minimizes the direct operating cost of hybrid-electric aircraft. This airspeed was then integrated into the RRT* algorithm used for path planning. Two scenarios were used to validate and demonstrate the capabilities of the minimum DOC solutions and of the path planning method applied to hybrid-electric and electric aircraft. The applicability of the hybrid-electric aircraft in longer-range flights was shown in an international travel scenario and the cost advantage of the electric aircraft was showcased in a urban air mobility scenario.
\bibliographystyle{IEEEtran}
\bibliography{IEEEabrv,References}

\end{document}